\documentclass{aip-cp}

\usepackage[numbers]{natbib}
\usepackage{rotating}
\usepackage{graphicx}
\usepackage{tikz}
\newcommand{\preliminary}[2][3]{
\begin{tikzpicture}
	\node[anchor=south west,inner sep=0] (murks) at (0,0) {#2};
  \begin{scope}[x={(murks.south east)},y={(murks.north west)}]
    \node [rotate=30,scale=#1,text opacity=0.2]
    at (murks.center) {Preliminary};
  \end{scope}
\end{tikzpicture}
}

\begin{document}

\title{Strangeness Photoproduction At Extremely Forward Angles At The BGO-OD Experiment}
\author[aff1]{T. C. Jude\corref{cor1}\noteref{note1}}
\authornote[note1]{On behalf of the BGO-OD collaboration.}
\affil[aff1]{Physikalisches Institut, Universit\"at Bonn, Germany}
\corresp[cor1]{Corresponding author: jude@physik.uni-bonn.de}

\maketitle

\begin{abstract}
The BGO-OD experiment at the ELSA accelerator facility uses an energy tagged bremstrahlung photon beam to investigate the internal structure of the nucleon. 
The setup consists of a highly segmented BGO calorimeter surrounding the target, with a particle tracking magnetic spectrometer at forward angles.

BGO-OD is ideal for investigating low momentum transfer processes due to the acceptance and high momentum resolution at forward angles.
This enables the investigation of strangeness photoproduction where $t$-channel exchange mechanisms play a dominant role.
A detailed understanding of this low-momentum transfer region is also crucial for constraints in hypernuclei electroproduction,
and sensitive to any extended, molecular-like interactions that may contribute to reaction mechanisms. 

Progress in the study of $K^+\Lambda$(1405) differential cross sections and line shapes, 
$K^0$ photoproduction, and differential cross section measurements for $K^+\Lambda$ and $K^+\Sigma^0$ photoproduction at extremely forward angles is presented.
Opportunities for hypernuclei studies are also discussed. 

\end{abstract}

\section{INTRODUCTION}

Hadron spectroscopy has for many years been used to determine the interactions between the partons of the nucleon, and the degrees of freedom afforded in non-perturbative QCD.
Despite the wealth of data for both the pion and photoproduction of many hadronic states, there remain many ``missing resonances'' which are predicted by Constituent Quark Models (CQM)
but are not observed experimentally~\cite{klempt10}.  Moreover, some of the lowest observed states are not described satisfactorily.  The pattern of the mass and parity of the Roper resonance
($N$(1440)1/2$^+$) and the $N$(1535)1/2$^-$ for example, where the state above ground state would be expected to have negative parity, is hard to reconcile with a CQM of ``dressed'' quarks
in a mutually generated potential, irrespective of the shape of this potential.  The \newline $\Lambda$(1405) - $N^*$(1535) mass ordering (despite the $\Lambda$(1405) being a $uds$ singlet state), 
and the mass between the $\Lambda$(1405) and its spin-orbit partner, $\Lambda$(1520), are also difficult to interpret within a CQM framework.

In the ``heavy'' charmed quark sector, the recently discovered pentaquark states, $P_c$(4450)$^+$ and $P_c$(4380)$^+$~\cite{aaij15} are the first unambiguous indications of baryons of at least five constituent quarks.
Since the conception of the quark model, there has been discussion of the possibility of hadrons of more than three constituent quarks~\cite{gellmann64, jaffe77, strottmann79},
 and due to the proximity of the chiral symmetry breaking scale to the nucleon mass, 
it is possible that light mesons may interact as elementary objects, giving rise to molecular systems and meson rescattering effects near thresholds~\cite{manohar84, glozman96}.
It is still an open as to question whether the pentaquark states are bound five quark systems, or have a molecular-like composition to some extent.
The model of Wu \textit{et al.}~\cite{wu10b}, for example, successfully describes these as meson-baryon dynamically generated states.  Similarly in the meson sector, the $X$(3872) lies close to the $D^0\bar{D}^{0*}$ threshold and has
also been described as a molecular state (see for example Ref.~\cite{tornqvist04}).  

In the ``light'' strange quark sector, models including molecular-like meson-baryon interactions as additional degrees of freedom have had improved success in describing observed states
 ~\cite{dalitz67, siegel88, kaiser97, recio04, lutz04}.  The $\Lambda$(1405) appears to be dynamically generated from meson-baryon interactions to some extent~\cite{nacher03},
which is also supported by recent LQCD calculations~\cite{hall15}.  
Models including vector meson-baryon interactions have predicted further dynamically generated states, for example states at 2~GeV with $J^P = 1/2^-$ and $3/2^-$~\cite{gonzalez09, sarker10, oset10},
which may have been observed in $K^0\Sigma^+$ photoproduction~\cite{ralf, ewald14} at the $K^*Y$ thresholds.  

The BGO-OD experiment is ideally suited to study phenomena from hadronic reactions in a low momentum exchange region, where extended, molecular-like structure may manifest.
The extremely forward charged particle acceptance, complemented by neutral particle reconstruction over a central region allows complicated final states to be reconstructed, 
enabling the study of phenomena from molecular or exotic structure in the strange quark sector.

\section{EXPERIMENTAL SETUP}

\begin{figure}[t]
  \centerline{\includegraphics[width=0.5\linewidth]{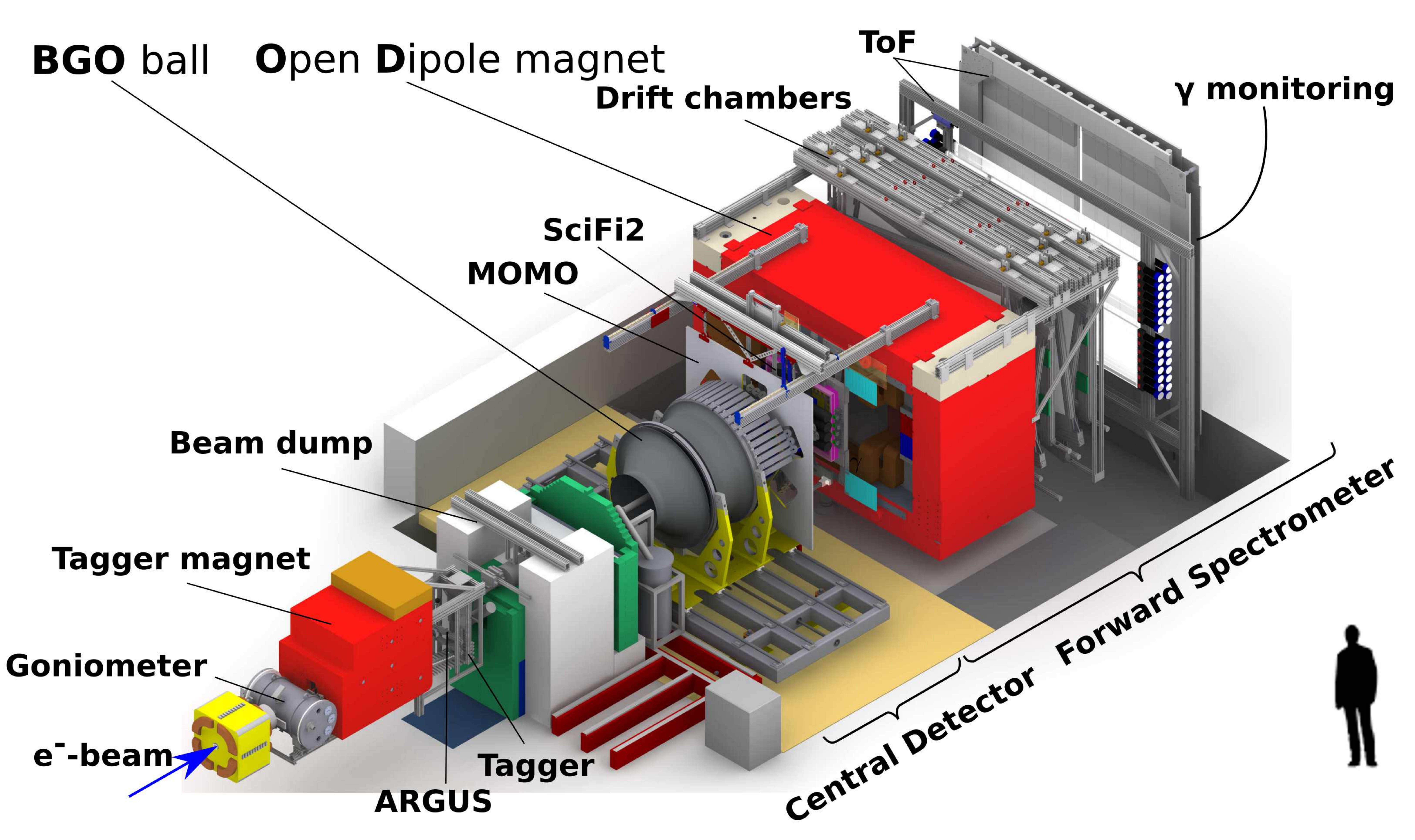}\includegraphics[width=0.5\linewidth]{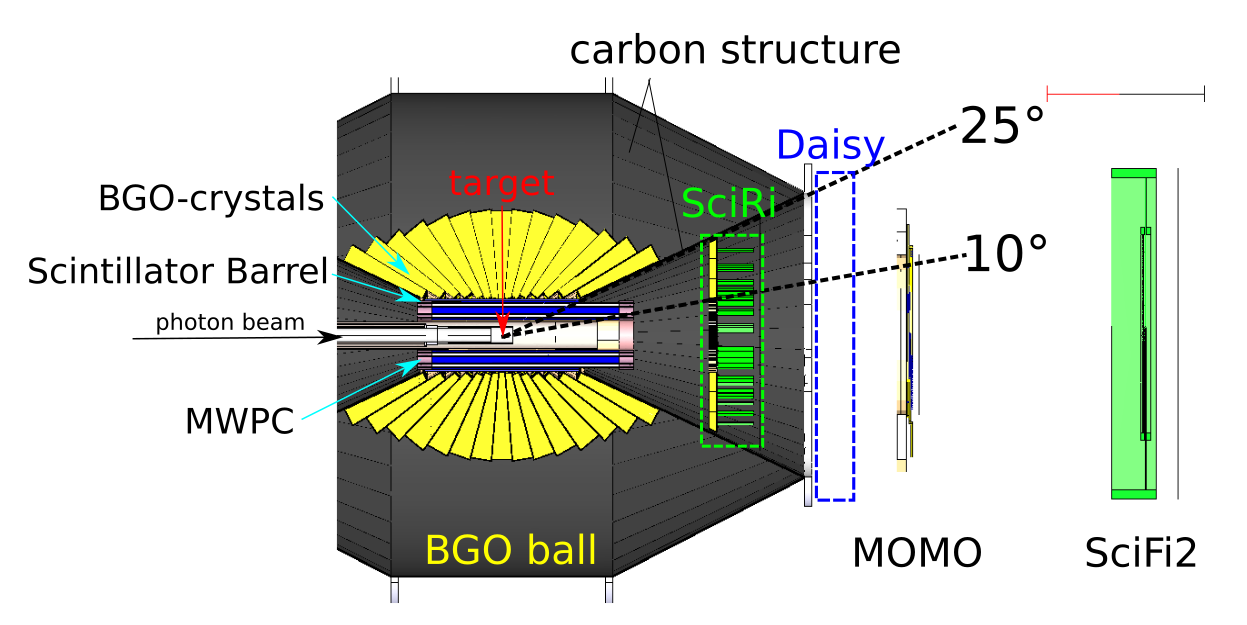}}
  \caption{The BGO-OD experiment at the ELSA facility, Bonn, Germany}\label{fig:bgood}
\end{figure}

BGO-OD (Fig.~\ref{fig:bgood}) is a fixed target experiment using real, energy tagged photon beams at the ELSA electron accelerator facility~\cite{hillert06}.
An electron beam up to 3.2~GeV is incident upon a thin metal radiator
to produce bremsstrahlung photons, with linearly and circularly polarised beams both available.

BGO-OD is composed of two distinct parts: a forward spectrometer ($\theta < 12^0$) for charged particle identification, and a central calorimeter region ($\theta = 25-155^0$) ideal for neutral meson reconstruction.
A plastic scintillating detector (SciRi) covers the small acceptance gap between these.  

\begin{figure}[h]
  \includegraphics[width=\linewidth]{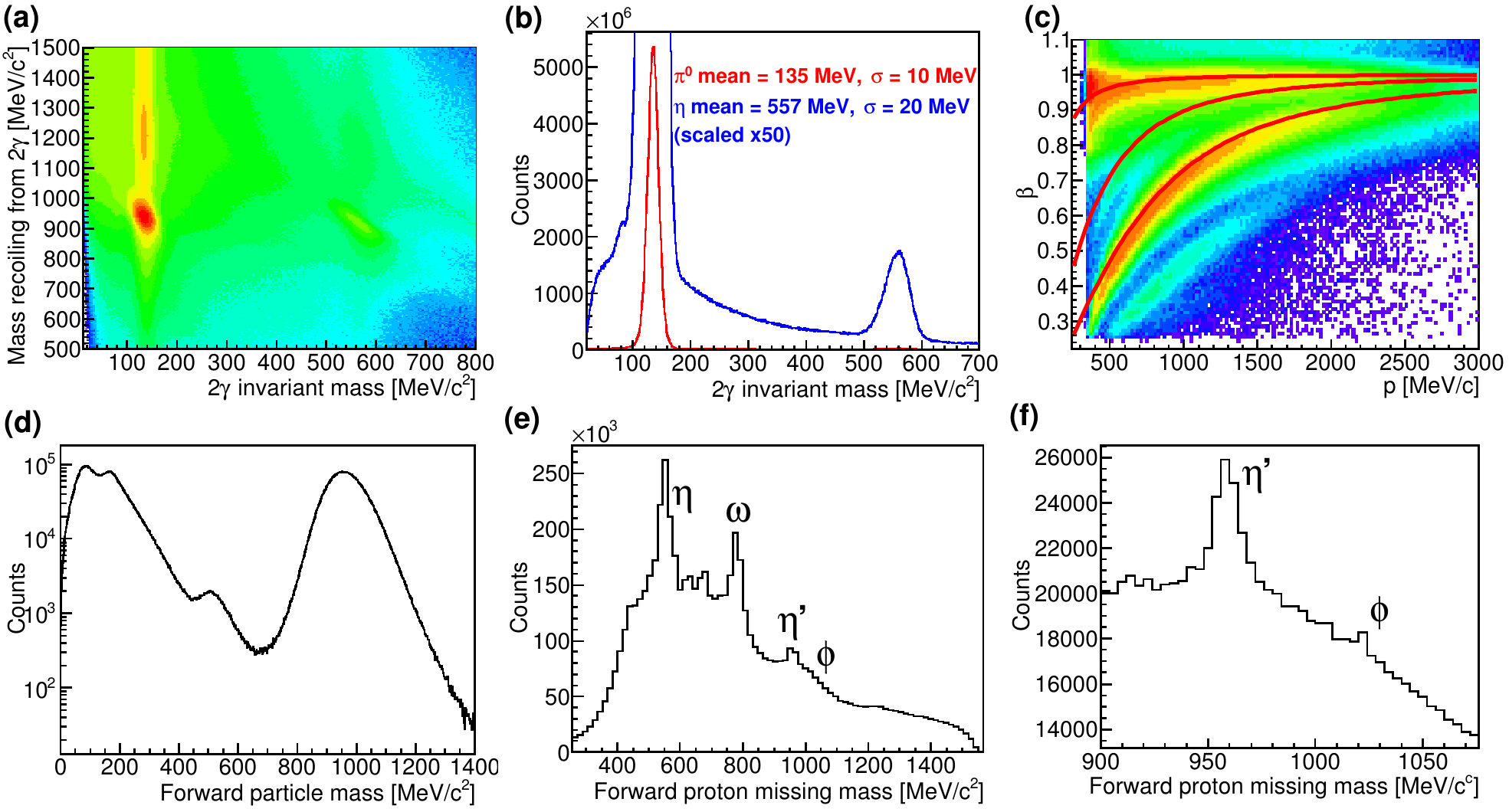}
  \caption{Neutral meson reconstruction in the BGO calorimeter (a,b) and  charged particle identification in the forward spectrometer (c-f).
(a) The missing mass recoiling from two photons in the final state versus the invariant mass of the two photons.  
Peaks corresponding to $\pi^0p$ and $\eta p$ final states are visible.  
(b) The invariant mass of two photons for events where the missing mass is within two sigma of the proton mass.  
Mean and sigma of a Gaussian fit are inset for $\pi^0$ and $\eta$ mesons.
(c) Forward charged particle $\beta$ (derived from time of flight) versus particle momenta.
Red lines indicate the expected loci of charged pions, kaons and protons.  (d) Particle mass calculated from $\beta$ and momentum.
(e) Missing mass from forward protons.  Peaks corresponding to single meson final states are labelled inset.  
(f) The same as (e) but with a different scale to observe the small peak corresponding to the $\phi p$ final state.}\label{fig:2gamma}
\end{figure}

The target is surrounded by an MWPC for charged particle track reconstruction.  
Surrounding this is a segmented cylinder of plastic scintillator material for charged particle identification via $\Delta E-E$.  
Outside of this is the BGO ball; a segmented calorimeter of 480 BGO crystals.
The BGO ball is ideal for the reconstruction of photon four-momenta via electromagnetic showers in the crystals.
The separate time readout per crystal, with a resolution of approximately 3~ns, enables clean identification of neutral meson decays (Fig.~\ref{fig:2gamma}(a,b)).

The forward spectrometer uses two scintillating fibre detectors (SciFi2 and MOMO) to track charged particles from the reaction vertex.
Particles proceed through the open dipole magnet, operating at a maximum field strength of 0.45~T.
Eight double layered drift chambers track particle trajectories after curvature in the magnetic field.
The momentum, $p$, is determined by the extent of deflection in the field, with a resolution of approximately $0.04p$.
Time Of Flight walls downstream from the drift chambers enable particle identification via the combination of momentum and $\beta$ (Fig.~\ref{fig:2gamma}(c-f)).

\begin{figure}[!h]
\preliminary{\includegraphics[width=0.5\linewidth]{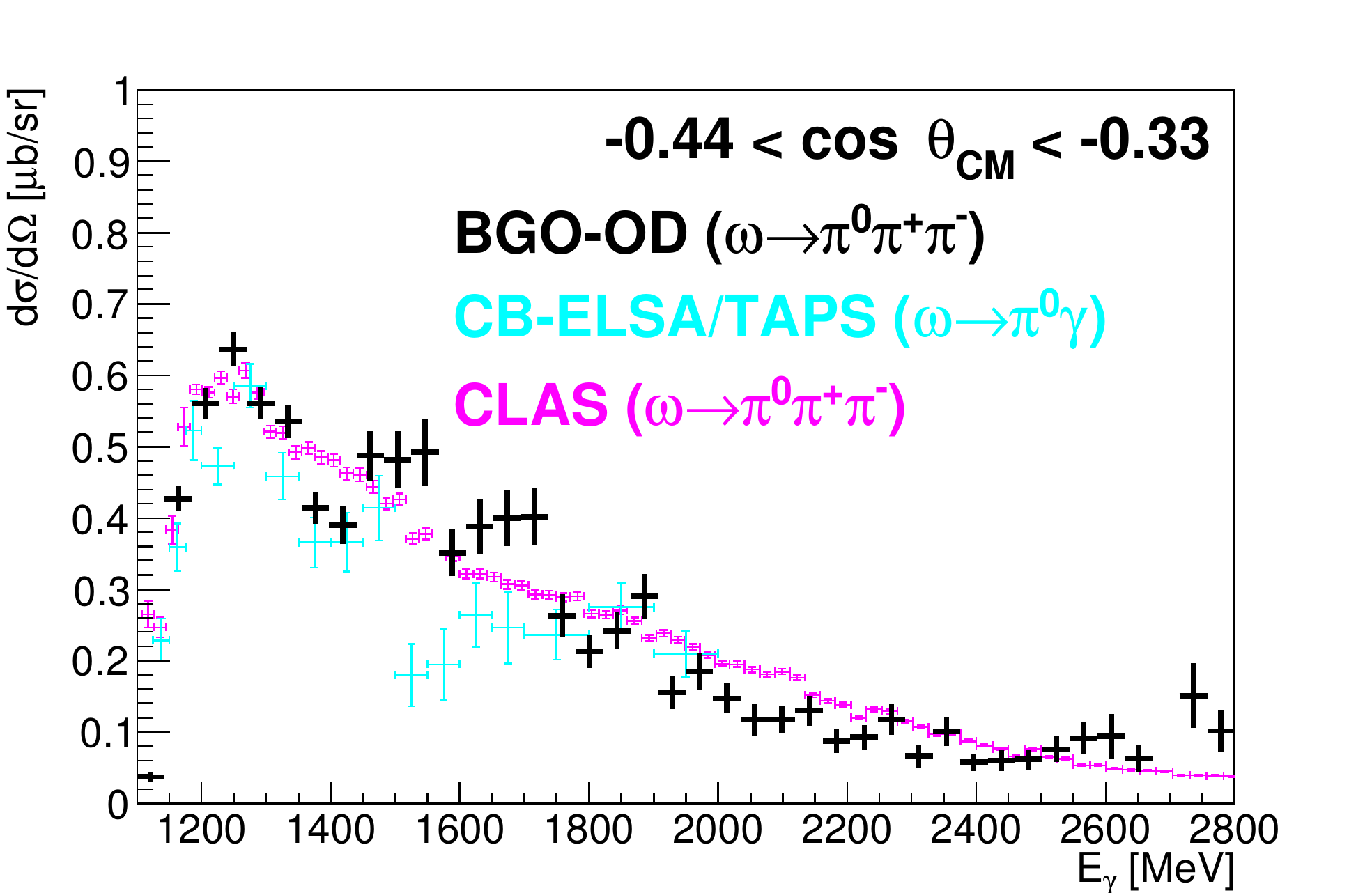}}\preliminary{\includegraphics[width=0.5\linewidth]{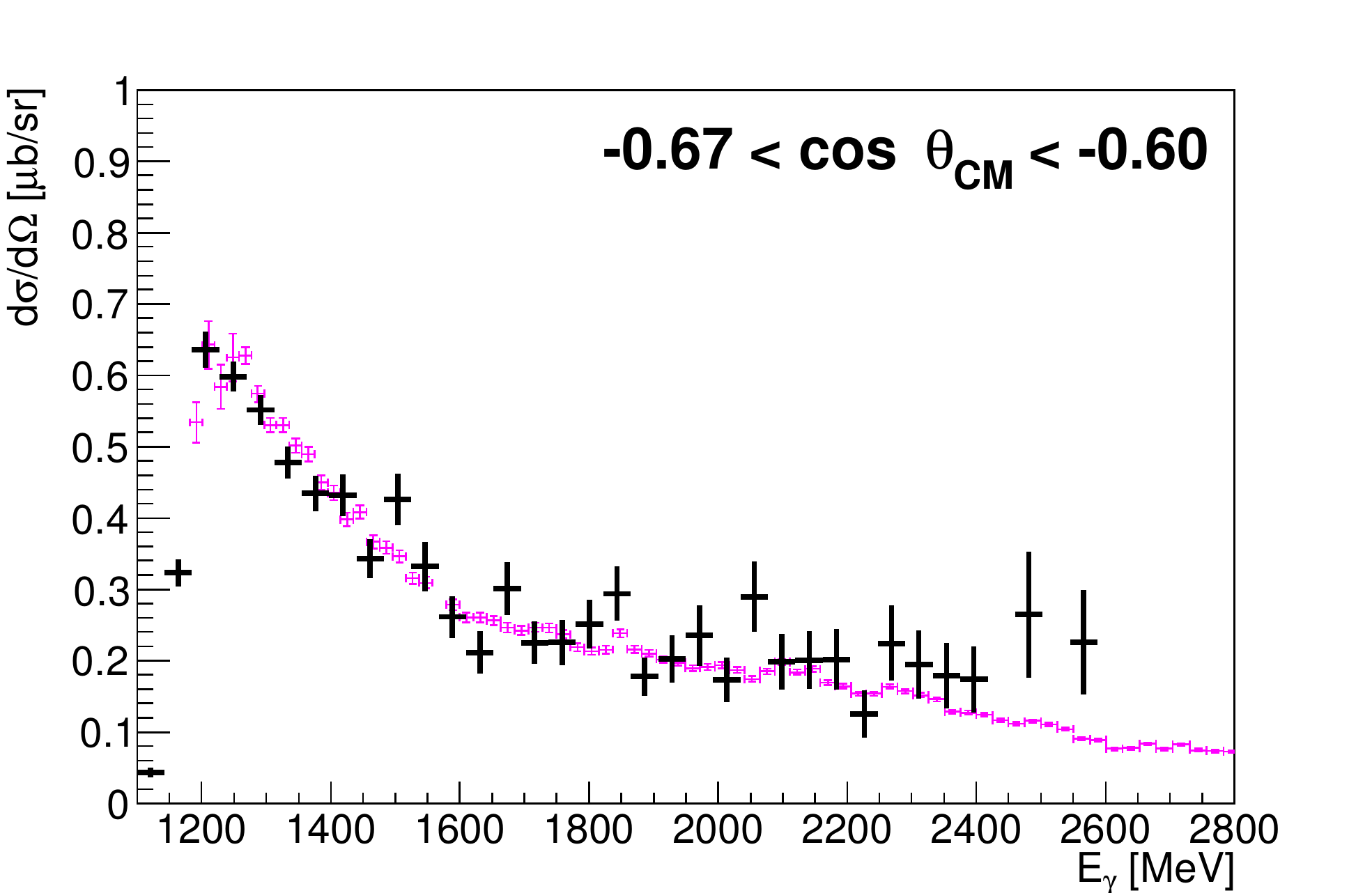}}
  \caption{$\gamma p \rightarrow \omega p$ differential cross sections for two centre of mass angle ranges, $\theta_{CM}$ (labelled inset), via the identification of the decay, $\omega\rightarrow \pi^0\pi^+\pi^-$.  
Preliminary BGO-OD data in black, previous data from F. Dietz \textit{ et al.} (CB-ELSA/TAPS collaboration)~\cite{dietz15} in cyan, previous data from M. Williams \textit{ et al.} (CLAS collaboration)~\cite{williams09} in magenta.}\label{fig:omega}
\end{figure}

Numerous hadronic final states have been identified, and cross sections calculated to act as a bench mark and to determine systematic uncertainties in analyses.
Figure~\ref{fig:omega} is a preliminary example of the $\gamma p \rightarrow \omega p$ differential cross section, identified via the mixed final state; $\omega\rightarrow \pi^0\pi^+\pi^-$.  
From a comparitively limited dataset, there is good agreement with existing data shown in the figure. 

\section{$\Lambda$(1405) LINE SHAPE AND DIFFERENTIAL CROSS  SECTIONS, AND ACCESS TO HIGHER LYING HYPERONS}

CQMs have very limited success in describing $\Lambda^*$ and $\Sigma^*$ states, 
with models including additional degrees of freedom from molecular-like meson-baryon interactions having far better agreement.
Of the 26 $\Sigma^*$ listed in the Particle Data Group review, 16 are listed as an existance as poor to fair, with little progression for the last 30 years~\cite{pdg}.

\begin{figure}[t]
  \includegraphics[width=\linewidth]{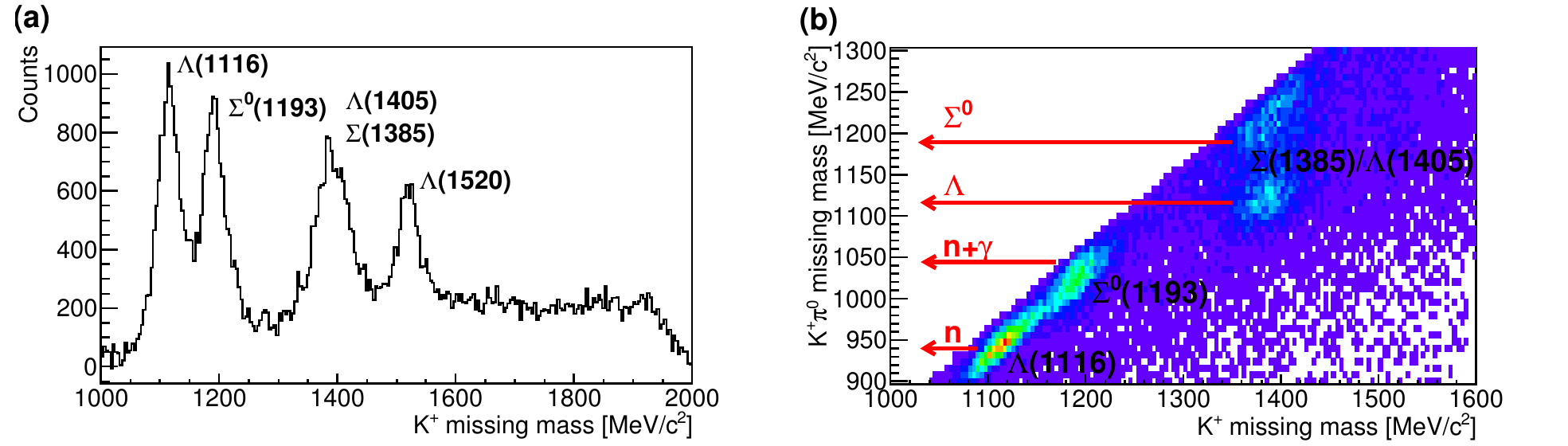}
  \caption{$K^+Y^{*}$ identification in the forward spectrometer.
 (a) Missing mass from forward $K^{+}$.  
Hyperons (labelled inset) are observed in this low momentum exchange region up to a maximum mass of approximately 2~GeV/c$^2$.
(b) The missing mass from the $\pi^0 K^+$ system is plotted against the missing mass from the $K^+$.
The red and black labels indicate the expected mass missing from the $\pi^0 K^+$ system and the mass missing from the $K^+$ system respectively.}\label{fig:hyperons}
\end{figure}

Figure \ref{fig:hyperons}(a) shows the missing mass from the detection of $K^+$ in the BGO-OD forward spectrometer.  
This extremely low momentum transfer region provides a unique window into the potential molecular-like structure of hyperons.  Peaks corresponding to the ground state $\Lambda$ and $\Sigma^0$ are clear,
and higher lying hyperons $\Lambda$(1405), $\Sigma$(1385), $\Lambda$(1520) are visible with no further selection criteria.
Hyperons overlapping in mass can be separated from the detection of $\pi^0\rightarrow \gamma\gamma$ in the BGO.  
Figure~\ref{fig:hyperons}(b) shows the missing mass recoiling from the $K^+\pi^0$ system versus the  missing mass recoiling from the $K^+$.
For the decays $\Lambda$(1405)$\rightarrow \pi^0\Sigma^0$ and $\Sigma$(1385)$\rightarrow \pi^0\Lambda$, for example, the missing mass from the $K^+\pi^0$ system are $\Sigma^0$ and $\Lambda$ respectively.
There is ongoing analysis using this method to determine the line shape and differential cross section for $\Lambda$(1405)$\rightarrow \pi^0\Sigma^0$ at forward angles.
This low momentum-exchange region is crucial in determining the underlying dynamics of the $\Lambda$(1405) and complementary to existing data, notably, the datasets of all three decay modes from CLAS~\cite{moriya13}.

\begin{figure}[t]
  \preliminary{\includegraphics[trim={0cm 0cm 0 0},clip,width=\linewidth]{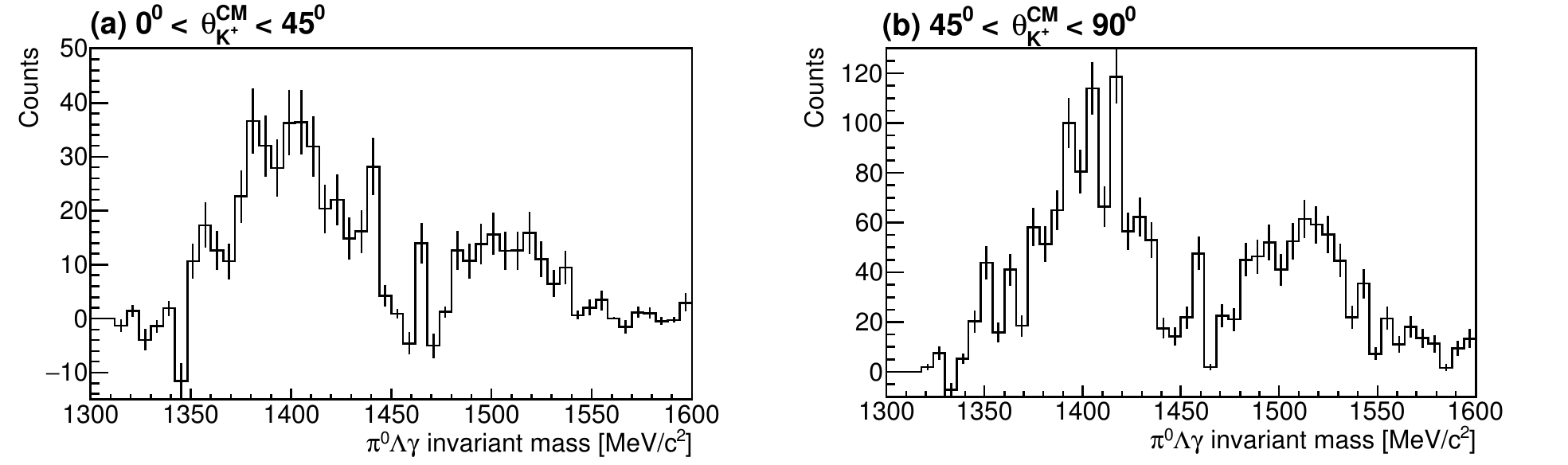}}
  \caption{Identification of $K^+\Lambda$(1405) and $K^+\Lambda$(1520) final states via identification of all decay particles from \newline $K^+ Y^*\rightarrow K^+\pi^0\Sigma^0\rightarrow\rightarrow K^+\gamma\gamma\gamma\pi^- p$~\cite{georg}.
The hyperon invariant mass is plotted for two different centre of mass bins, $\theta^{cm}_{K^+}$.}\label{fig:lambda1405invmass}
\end{figure}

The same decay mode can be identified over a broad angular range via the identification of all final state particles: $K^+ Y^*\rightarrow K^+\pi^0\Sigma^0\rightarrow\rightarrow K^+\gamma\gamma\gamma\pi^- p$,
and the application of a kinematic fit~\cite{georg}.  Figure \ref{fig:lambda1405invmass} shows the invariant mass for two different polar angle bins, where peaks corresponding to $\Lambda$(1405) and $\Lambda$(1520) are present.
This ongoing work will complement a limited data set, and aid in constraining models, such as the triangle singularity proposed by Wang \textit{et al.}~\cite{wang17}.

\section{FORWARD CROSS SECTIONS FOR $K^+\Lambda$ AND $K^+\Sigma^0$ PHOTOPRODUCTION}

The photoproduction of the ground state hyperons, $K^+\Lambda$ and $K^+\Sigma^0$ is poorly understood at forward angles.  
The paucity of data for $\cos\theta^{cm}_{K^+} > 0.9$ and discrepancies between data sets prevent the constraining of isobar models and partial wave analyses (shown, for example, in Refs.~\cite{skoupil18, bydzovsky12} and references therein).  
This is a vital region to understand reaction mechanisms in associated strangeness photoproduction, where forward peaked, $t$-channel mechanisms play dominant roles.
This is also crucial as an input for the modelling and data fitting of hypernuclei electroproduction:   
the low $Q^2$ required to ensure the $\Lambda$ remains bound to the nucleus means that the exchanged virtual photon is almost on-shell.

\begin{figure}[t]
 \preliminary{\includegraphics[trim={10cm 0 0 0},clip,width=0.5\linewidth]{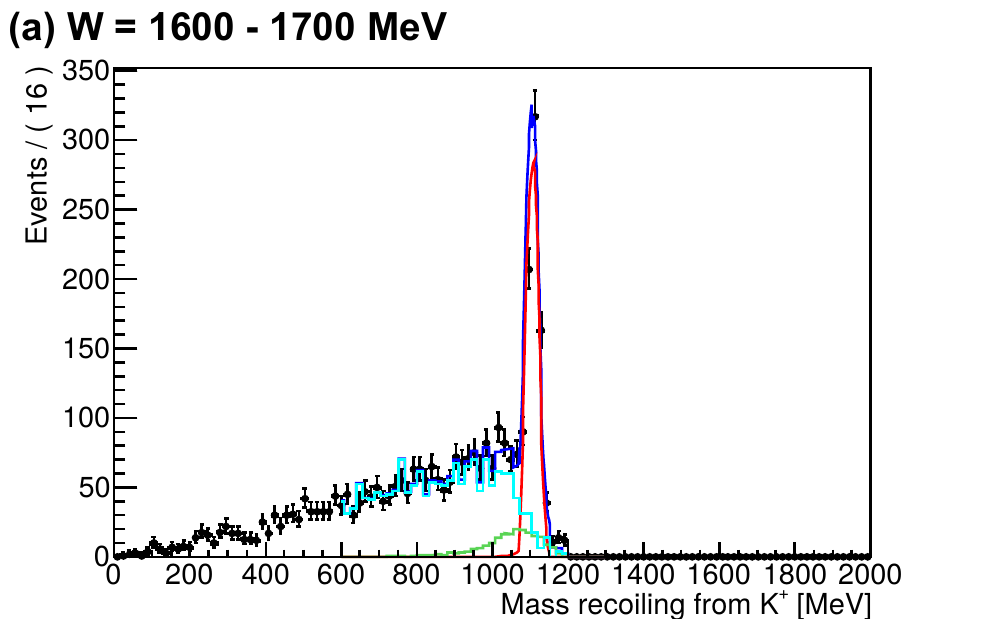}}\preliminary{\includegraphics[trim={10cm 0 0 0},clip,width=0.5\linewidth]{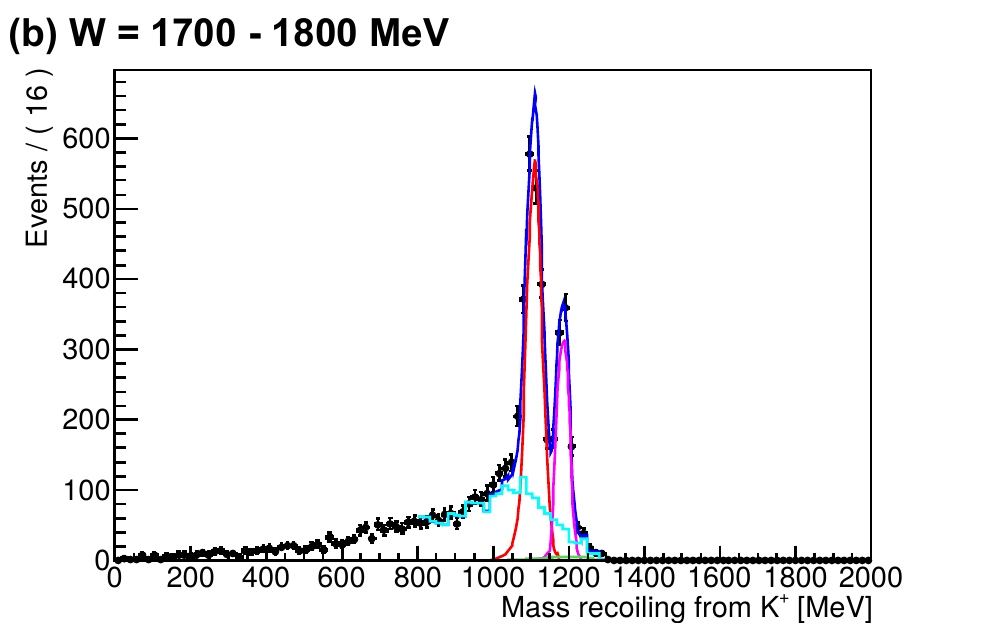}}
 \caption{Missing mass from forward $K^+$ for two centre of mass energy ranges.  
The black points are experimental data which have been fitted to using RooFit~\cite{roofit}: background from pair production in the beam (cyan line), simulated $K^+\Lambda$ (red line) and simulated $K^+\Sigma^0$ (magenta line).
The blue line is the sum from the three fit components.}\label{fig:mm}
\end{figure}

The BGO-OD forward spectrometer covers the centre of mass polar angle range of approximately 3-26$^0$, with an angular resolution better than 1$^0$.
Figure \ref{fig:mm} shows the missing mass from forward $K^+$ for two centre of mass intervals.  Peaks corresponding to $\Lambda$ and $\Sigma$ (only for $W$ above threshold) are clear.
The background tail to lower mass originates from pair production in the beam which are deflected in the OD magnet.  
Careful timing criteria remove most of this, which originates from adjacent electron bunches in time to the triggered hadronic event.  
The remainder is fitted to by an equivalent analysis of negatively charged particles in the forward spectrometer, providing a distribution of electron/positron background.
This is combined with spectra from simulated data to fit to signal and background using RooFit~\cite{roofit}, and to separate $K^+\Lambda$ and $K^+\Sigma^0$ events.  

\begin{figure}[t]
\preliminary{\includegraphics[clip,width=\linewidth]{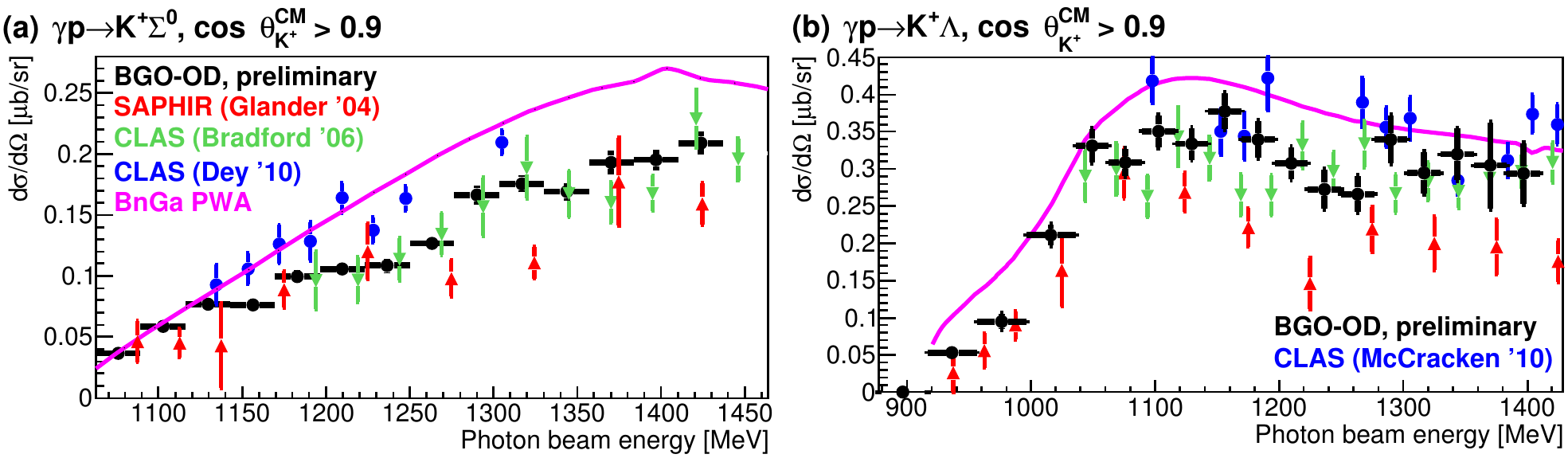}}
 \caption{Preliminary forward differential cross sections for (a) $K^+\Sigma^{0}$ and (b) $K^+\Lambda$ photoproduction.  
Bold black points: preliminary BGO-OD data,  vertical errors bars are the statistical error, horizontal error bars are the width of each energy interval.  Red triangles: data from Glander \textit{et al.}~\cite{glander04}, 
inverted green triangles: data from Bradford \textit{et al.}~\cite{bradford06},
blue circles: data from Dey \textit{et al.}~\cite{dey10} and McCracken \textit{et al.}~\cite{mccracken10} for $K^+\Sigma^0$ and $K^+\Lambda$ respectively, magenta line: solution from the Bonn-Gatchina Partial Wave Analysis~\cite{bnga}.}\label{fig:cs}
\end{figure}

Figure.~\ref{fig:cs} shows preliminary differential cross sections for $\gamma p \rightarrow K^+\Lambda$ and  $K^+\Sigma^0$ for $\cos^{CM}_{K^+} > 0.9$.  
This high statistics data set shows good agreement with existing data for $K^+\Sigma^0$, and is able to resolve long standing discrepancies in the $K^+\Lambda$ data sets.
It should be noted that the Bonn-Gatchina partial wave analysis~\cite{bnga} is constrained by the CLAS data sets only.
The high angular resolution of the forward spectrometer permits an unprecented detailed description of the cross section at forward angles, which will be a vital constraint for hypernuclei electroproduction.
There is ongoing analyses to measure both cross sections in intervals of approximately 0.02 $\cos^{CM}_{K^+}$, and it is anticipated the data will extend to a photon beam energy of 1.5~GeV,
with a 1.5 increase in statistics.

\section{NEUTRAL KAON PHOTOPRODUCTION}

The BGO-OD experiment is uniquely suited for $K^0_S$ identification, as the mixed charged final state reconstruction allows identification via both $K^0_S\rightarrow \pi^0\pi^0$ (neutral decay) and $K^0_S\rightarrow\pi^+\pi^-$ (charged decay).
High statistics data with a liquid hydrogen target has been taken with linearly polarised photons over the $K^*$ threshold region.  
Differential cross sections and beam asymmetry measurements will resolve reaction mechanims where a cusp-like structure was observed in previous data~\cite{ralf,ewald14}.  
As discussed in the introduction, similarly to the model of Wu \textit{et al.,}~\cite{wu10b} which described the pentaquark states, $P_c(4450)^+$ and $P_c(4380)^+$ as dynamically generated through rescattering effects, 
the model of Oset~\textit {et al.}~\cite{oset10}, has been fitted to the cusp in $K^0\Sigma^+$ cross section data~\cite{ralf,ewald14}, predicting dynamically generated meson-baryon states in the ``light'' strange quark sector.

\begin{figure}[h]
\preliminary{\includegraphics[trim={0cm 0 0 0},clip,width=\linewidth]{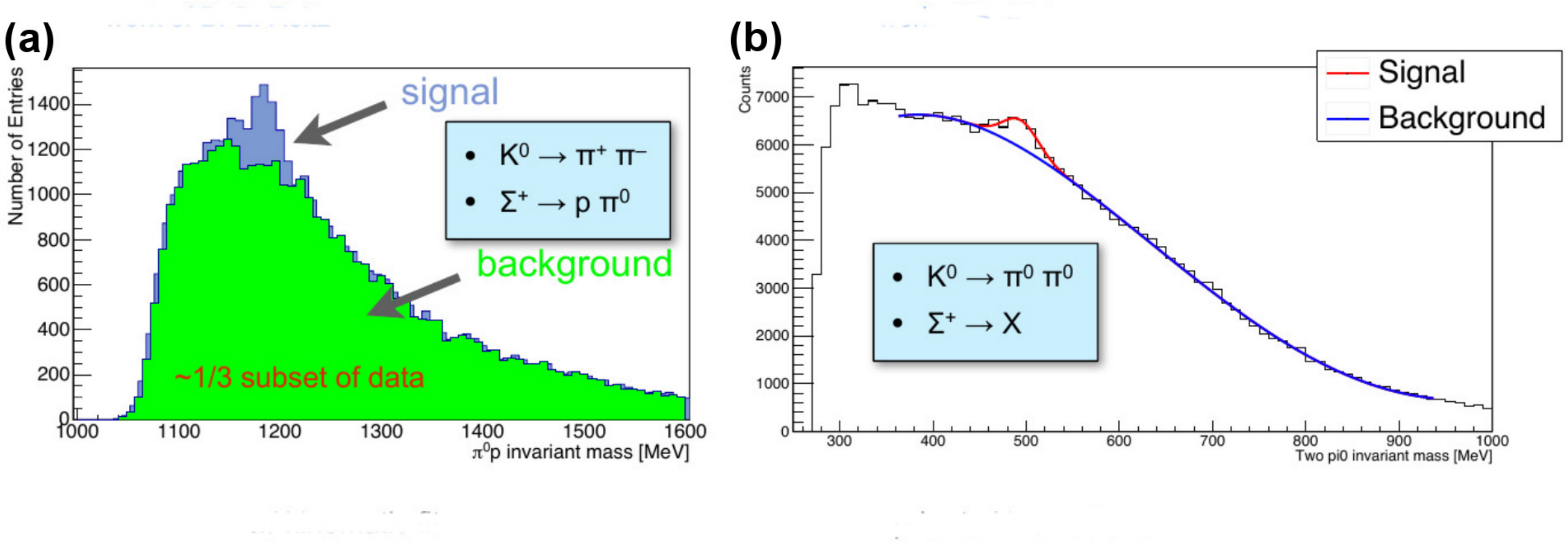}}
 \caption{$\gamma p \rightarrow K^0_S\Sigma^+$ reconstruction.  (a) Identification via the charged decay, $K^0_S\rightarrow \pi^+\pi^-$.  
The $\Sigma^+$  invariant mass peak is shown in blue, with background contributions shown in green~\cite{bjoern}.  
(b) Identification via the neutral decay, $K^0_S\rightarrow \pi^0\pi^0$.  
A polynomial and Gaussian function have been fitted to background and $K^0_S$ invariant mass signal respectively~\cite{stefan}.}\label{fig:k0}
\end{figure}

Figure~\ref{fig:k0}(a) shows preliminary invariant mass of the $2\pi^0$ system, with a peak at the $K^0$ mass.  Simulated data was used to describe background from other multi-pion final states.  
Figure~\ref{fig:k0}(b) shows the $\Sigma^+$ invariant mass peak via the identification of the  charged $K^0_S$ decay.
Both analyses are prior to a kinematic fit, which is currently being implemented.  The signal from the charged decay will also be improved by identifying the detached decay vertex using the MWPC ($c\tau_{K^0_S} \approx 2.7$~cm).

\begin{figure}[h]
\preliminary{\includegraphics[trim={0cm 13.1cm 0 0},clip,width=0.5\linewidth]{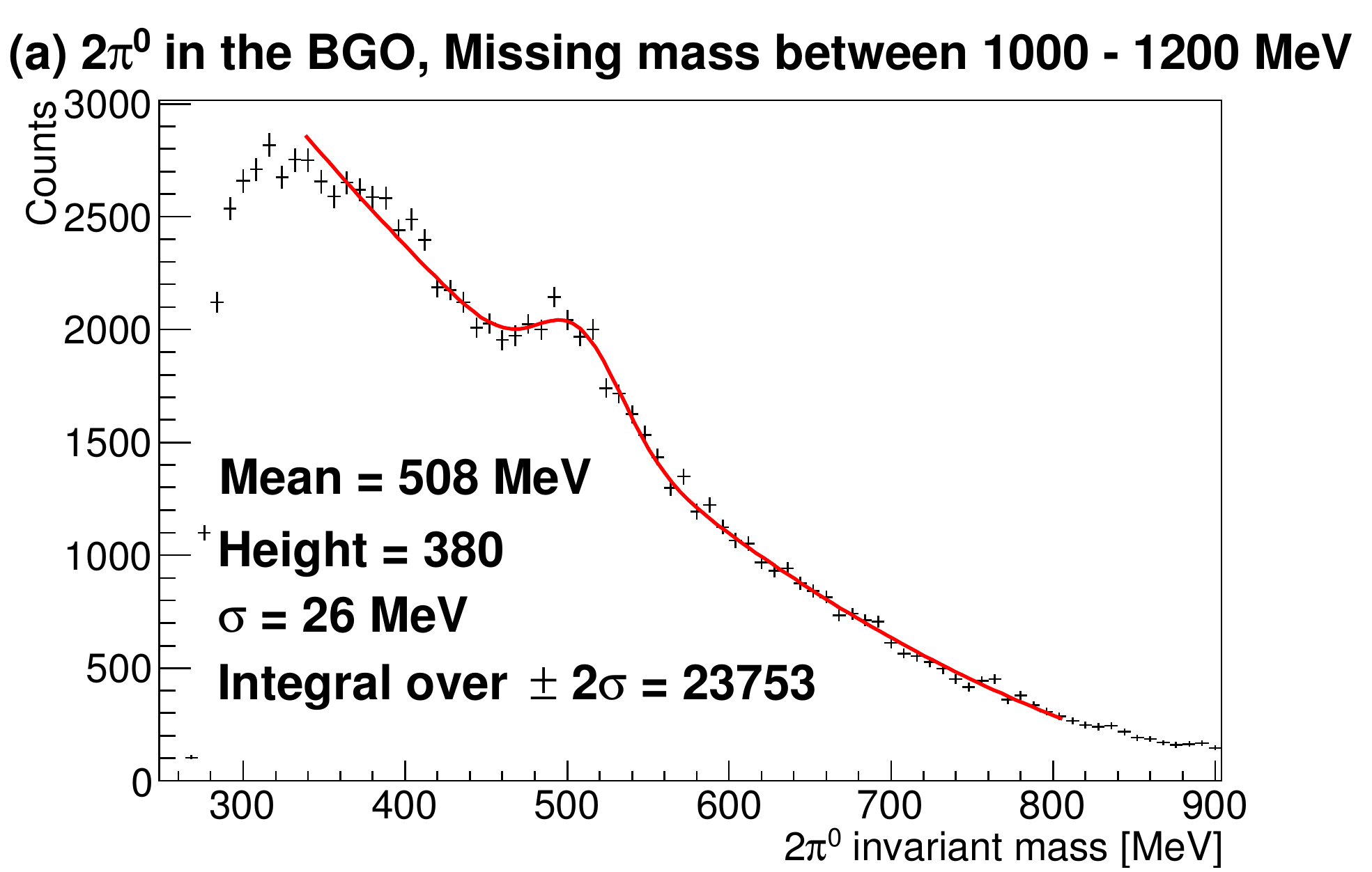}}\preliminary{\includegraphics[trim={0cm 0 0 13cm},clip,width=0.5\linewidth]{K0_InvMassOnly-eps-converted-to.pdf}}
 \caption{$K^0_S\Lambda$ and $K^0_S\Sigma^0$ reconstruction using a deuterium (neutron) target.  (a) Two $\pi^0$ invariant mass. 
A polynomial and Gaussian function are fitted to describe background and the mass peak from the $K^0_S$ respectively, with the Gaussian function parameters inset.  
(b)  Equivalent to (a) but with an additional proton identified in the forward spectrometer from the $\Sigma^0\rightarrow p \pi^-$ decay.}\label{fig:k0neutron}
\end{figure}

The model of Oset \textit{et al.}~\cite{oset10} predicts that the states that interfer destructively to produce a cusp in $K^0_S\Sigma^+$ photoproduction will interfer constructively in $K^0_S\Sigma^0$ to produce a peak-like structure at the $K^*$ thresholds.
The study of this channel using a deuterium target will therefore act as a ``smoking gun'' for molecular-like structure in the light quark sector.
Figure~\ref{fig:k0neutron} shows preliminary $K^0_S$ identification from a small commissioning data set using a deuterium target.  A peak in the $2\pi^0$ invariant mass spectrum is clearly seen above background.  
Figure~\ref{fig:k0neutron}(b) additionally requires a proton in the forward spectrometer from $\Sigma^0\rightarrow p\pi^-$ and for the missing mass from the $2\pi^0p$ system to be consistent with a $\pi^-$ mass.  
This improves the signal to background ratio, albeit at the loss of statistics.  An equivalent analysis using liquid hydrogen target did not produce a signal.
Analyses using both real and simulated data has demonstrated the separation of $K^0\Sigma^0$ and $K^0\Lambda$ via the identification of the photon from the $\Sigma^0\rightarrow \gamma\Lambda$ decay.
A high statistics data set using a deuterium target was taken this year, and analyses is ongoing~\cite{katrin}.

\section{OPPORTUNIITES TO STUDY THE YN INTERACTION AND HYPERNUCLEI}

Hypernuclei, where baryons of non-zero strangeness are bound to a nucleus, provide a natural laboratory to probe hyperon-nucleon (YN) interactions.
YN or YY (or three body YNN, YYN) interactions are very poorly constrained compared to NN interactions.
Studying these is vital in order to develop an SU(3)$_\mathrm{flavour}$ symmetric understanding of baryon interactions.
Additionally, the description of many astrophysical phenomena depend strongly on YN interactions, for example, the \textit{Hyperon Puzzle} in describing the equation of state of neutron stars (see for example, Ref. \cite{gal16} and references therein).

\begin{figure}[h]
\preliminary{\includegraphics[clip,width=\linewidth]{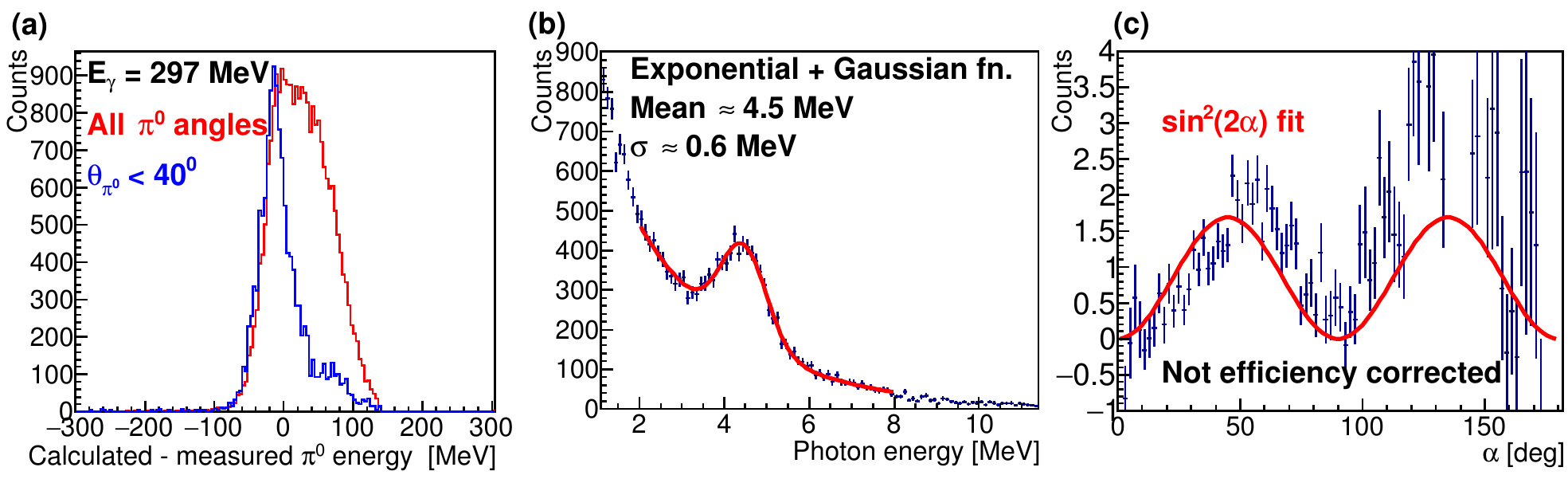}}
 \caption{Identification of coherent $\pi^0$ photoproduction off carbon, as a demonstration of potential future opportunities in hypernuclei research.  
(a) Difference between calculated and measured $\pi^0$ energy with a beam energy of 297~MeV for all angles and at polar angles smaller than 40$^0$.  A peak at zero indicates coherent events.
(b) Identification of the 4.4~MeV decay $\gamma$ from an excited carbon nuclei.  An approximate fit of a exponential to background and Gaussian function to signal yields the mean and sigma inset.
(c) Angular decay distribution of the decay $\gamma$ with a $\sin^2(2\alpha)$ fit (see text for details).  }\label{fig:coherentpion}
\end{figure}

The BGO-OD experiment, using a real photon beam, may be able to make complementary measurements to existing hypernuclei facilities.  
The energy resolution of the incident photon beam will not permit measurements of hypernuclei binding energies, 
however the high momentum resolution of $K^+$ at forward angles, and a calorimeter to identify both charged and neutral decay particles provide unique opportunities.  
From simple kinematics, the BGO-OD is ideal: an incident photon beam between 1400 to 1600 MeV, and a $K^+$ produced at a polar angle smaller than 10$^0$ provides a recoil momentum to a $\Lambda$ of the order of 300 - 350 MeV/c,
which would be sufficient for some to remain within the Fermi surface and bind to the residual nucleus.  
Figure~\ref{fig:coherentpion} demonstrates a ``non-strange'' example, where coherent $\pi^0$ photoproduction was observed off a carbon target.
Figure~\ref{fig:coherentpion}(a) shows the calculated $\pi^0$ energy from the polar angle and beam energy, with the measured energy subtracted.  The peak at zero for forward $\pi^0$ indicates coherent production.  
For further verification, the 4.4~MeV decay photon from the full reaction: \newline$\gamma + ^{12}C \rightarrow \pi^0 +  ^{12}C^* \rightarrow \pi^0 + ^{12}C + \gamma'$ was identified (shown in fig.~\ref{fig:coherentpion}(b)).
This decay corresponds to a pure E2 transition, $J^\pi: 2^+\rightarrow 0^+$, with a $\sin^2(2\alpha)$ distribution where $\alpha$ is the angle between the recoiling nucleus and decay gamma.  
Figure~\ref{fig:coherentpion}(c) shows this angular distribution with a $\sin^2(2\alpha)$ fit.  There is good agreement when accounting that the data is not efficiency corrected.
\newpage
These techniques are a first step in demonstrating that hypernuclei research may be feasible.  
BGO-OD may provide opportunites in measuring angular distributions of hypernuclei photoproduction~\cite{mainz}, which has some theoretical support~\cite{petrtalk}, 
however it is not clear how useful the data are without a separation of different states.
There may also be opportunities to use targets which are spoiled by intense electron beams, or to measure hypernuclei lifetimes with the development, for example, of an active target~\cite{mainz}.
\bibliographystyle{aipnum-cp}%


\end{document}